# Noise Pollution and Household Sustainability: An Economic Approach

Yi Fan[i]

**Summary**

Examining the economic impact of noise pollution from a lens of household is a burgeoning field in the study of environmental sustainability. Economics studies cover the source, measure, consequence of noise pollution, as well as the econometric methods used to identify the causal impact of noise pollution on socioeconomic welfare. There are broadly four major noise origins along with the industrial growth and urban development, which are airport, railway, urban traffic, and neighborhood. Four general kinds of measures or data sources are used in economics studies to capture the noise variations, namely, proximity to noise origins, real-time noise monitor records, household surveys, and administrative records on noise complaints. The socioeconomic consequences of noise pollution span from physical or mental health to happiness, violence and suicide, housing market capitalization, and inequality. In economics studies, generally three types of econometric methods are used to identify causal impact of noise pollution on the household's welfare, which are instrumental variable estimation, difference-in-difference estimation, randomized and quasi-natural experiments. The causal impact of noise pollution on household's socioeconomic welfare derived from economics studies can help guide policy efforts in allocating resources for noise elimination and conduct cost-benefit analysis. The economics research contributes to the general noise research from both conceptual and methodological perspectives: It expands the scope of research from sound-poof technology or site layout planning to human welfare, and endeavors to isolate the causal impact of noise pollution from other confounding factors. Future studies are warranted along the lines of environmental injustice of noise pollution and socioeconomic consequences in less developed countries when the data become more available.

**Keywords**: noise pollution, household sustainability, economic approach, source of noise, noise measure, socioeconomic consequences

## Noise Pollution and Household Sustainability

Air and water are widely accepted as public goods, the pollution of which is considered as a Tragedy of the Commons by economists. However, another type of public resource—quiet environment—is often overlooked, possibly due to an oversimplified perception that noise is an "*unavoidable by-product of desirable technologies, transportation, and recreational activities*" (Hammer et al., 2018), and complicated by the difficulties in capturing the real-time noise levels.

Nevertheless, noise pollution has substantial impact on the sustainable growth of households, which are the fundamental elements of a society. Noise can cause long-term physical and mental health issues, such as cardiovascular problems (Fan and Weinhold, 2022), performance degradation (Zhang et al., 2021), and stress process (Evans et al., 1998). At the aggregate level, it has ramifications in various socioeconomic aspects, such as increase in the violent crime rate (Hener, 2022), decrease of property values in the neighborhood (Fan et al., 2021; Wang et al., 2023), and even declining fertility and raised mortality of killer whale (Taylor and Mayer, 2023). The World Health Organization (WHO) reports more than 1.6 million healthy life-years loss per year from traffic-related noise in Western Europe only (WHO, 2011). Traffic noise is merely one type of noise pollution. The sources of noise are ubiquitous: from the airport at the city fringe to the urban traffic in the Central Business District (CBD), from the cross-country rails to the renovation activities in the neighborhood.

Engineer literature has long debates on the sound-proof technology and site layout planning as solutions to noise pollution (Murphy and King, 2010; Hammad et al., 2017; Yu et al., 2020; Hasegawa and Lau, 2022a,b). However, it is only until recent decades that increasingly more economists pay attention to capitalizing noise pollution in the market and estimating its socioeconomic implications. One possible reason arises from increasing awareness of the adverse effects of noise pollution on welfare. On the other hand, the growing access to big data provides an arena for econometric analysis to generate reliable estimates to draw conclusions. This paper summarizes and discusses the source, measure, and consequence of noise pollution, from the lens of households and using an economic approach. This chapter aims to pave the foundation for the burgeoning economic studies on the impact of noise pollution on the survival and thriving of households, growing in a sustainable manner.

It starts with discussing the major sources of noise pollution. The early economics literature examines the impact of noise from airports—which is a symbol of city development—on health of residents and property values in the vicinity (Bugliarello et al., 1976; Nelson, 1979). It then extends to cross-country railway (Diao et al., 2016; Trojanek, 2023) and within-country road traffic (Babisch et al., 2001; Kim et al., 2019; Forouhid et al., 2023; Pradeep and Nagendra, 2024). Studies in the early 21$^{st}$ century go into more granular level of neighborhoods, investigating noise from anthropogenic activities (Xu et al., 2020; Fan and Weinhold, 2022). Industrial sites are also taken into examination, such as wind farms (Jensen et al., 2014; Zou, 2020) and construction sites (Ning et al., 2019).

To measure the variations in noise from those sources, economists use distance to noise origins as proxy (Kuehnel and Moeckel, 2020; Wang et al., 2023; Razavi-Termeh et al., 2024), derive data from monitoring records (Babisch et al., 2001; Trojanek, 2023) household surveys (Weinhold, 2013; Kim et al., 2019), or administrative records of noise complaints (Fan et al., 2021). Pros and cons exist for each measure as a consequence of trade-off between convenience and accuracy in measuring the variation of noise.

The adverse effect of noise pollution covers various welfare aspects of individuals and households. It ranges from physical to mental diseases (Cameron et al., 1972; Miki et al., 1998; Babisch et al., 2001; Münzel et al., 2021), from declined happiness to increasing crime and suicide rates (Van Praag and Baarsma, 2005; Weinhold, 2013; Zou, 2020; Hener, 2022), as well as drop in property values in the vicinity of noise origins (Baranzini and Ramirez, 2005; Boes and Nuesch, 2011; Wang et al., 2023). Environmental injustice is also demonstrated in the noise context; some population or socioeconomic segments are more vulnerable to noise pollution than others, such as the lower-income and younger/older age groups (Zou, 2020; Bessone et al., 2021).

On top of expanding the scope of noise pollution research, economics literature also contributes to the noise literature from a methodological perspective by identifying the *causal* mechanism. Instead of presenting *association* analysis, economists are keen to isolate the *causality*: if and how much the change in outcome is caused by the variation in noise pollution. Econometric strategies, such as instrumental-variable estimation, Difference-in-Difference (DiD) estimation, randomized and quasi-natural experiments are used to overcome endogeneity concerns and exclude impacts from confounding factors (Winke, 2017; Boes and Nüesch, 2011; Bessone et al., 2021; Fan and Weinhold, 2022).

How can households achieve sustainable growth and eliminate the adverse impact of noise pollution? This paper discusses relevant policy endeavors, including vehicle tax, abatement investment, and Low Traffic Neighborhoods (Bergendahl, 1976; Friedt and Cohen, 2021b; Leach et al., 2024). Future endeavors are expected to be along the lines of expanding the knowledge on the socioeconomic outcomes of noise pollution in developing countries and among vulnerable populations.

## Source of Noise Pollution

### *Aircraft*

Early economics literature on noise pollution emerged around 1970s and mainly focused on the noise pollution in the vicinity of airport, which is a sign of industrial development and modern society (Jackson, 1971; Bugliarello et al., 1976; Nelson, 1979). The early batch of evidence is derived from pioneer countries benefited from the Industrial Revolution, such as the United Kingdom (Bugliarello et al., 1976) or European countries (Alenandre et al., 1980; Evans et al., 1998). The aircraft noise is concentrated around airports and is regular in frequency, following the flight schedules.

Early studies uncover the relationship between airplane noise and sleep disruption (Bugliarello et al., 1976) and psychophysiological stress among children (Evans et al., 1998); men have a significantly higher noised-induced health risk compared to women (Maschke et al., 2002). From the 2000s, the economics literature of aircraft noise focuses on capitalization in the housing market (Van Praag and Baarsma, 2005; Baranzini and Ramirez, 2005; Boes and Nuesch, 2011; Winke, 2017; Friedt and Cohen, 2021b). Friedt and Cohen (2021a) show that contour-based calculations severely underestimate aircraft-noise-pollution-induced losses incurred by homeowners.

### *Railway*

Another symbol of modern transport development, as well as source of noise pollution, is cross-country/city railway (Brainard et al., 2004; Diao et al., 2016; Trojanek, 2023). Those railways connect different cities in the region, such as the Eurostar connecting London, Paris, Amsterdam, and Brussels in the Europe, and the mass high-speed railway system in China. One example is Diao et al. (2016) who use the cessation of a nearly 80-year-old railway between Malaysia and Singapore as a quasi-natural experiment. They show the train noise externalities and find that it increases housing price near the railway by 3.5%. Another example is Trojanek (2023), who geocoded 16,247 apartments in Poznan, Poland. It is found that railway noise is the most significant noise source to influence apartment prices; an increase of 1dB above 55dB can induce approximately 1.8% decrease in apartment values in the vicinity of railways.

***Urban Traffic***

In addition to the airport and railway noise, economists also investigate the road and vehicle noise in the cities. Traffic noise from public and private vehicles are the major sources of urban traffic noise. The negative externalities of traffic noise span from developing countries like China, India, Brazil, Mexico, and Iran (Fiedler and Zannin, 2015; Khaiwal et al., 2016; Li, 2018; Forouhid et al., 2023; Mora-Quiñones et al., 2024; Pradeep and Nagendra, 2024) to developed world such as Sweden, Germany, South Korea, Ireland, and Singapore (Bergendahl, 1976; Babisch et al., 2001; Kim et al., 2019; Kuehnel and Moeckel, 2020; Basu et al., 2021; Fan et al., 2021).

In detail, such negative externalities include congestion, accidents, air pollution, and noise (Li, 2018). Traffic noise has significantly adverse effects on health, as measured by increasing chance of vascular dysfunction, cardiovascular diseases, and stress hormones (Babisch et al., 2001; Münzel et al., 2021). It can also translate into lower housing rents by capitalizing the noise pollution into living close to roads (Wang et al., 2023). Policy wise, vehicle taxes and congestion charges are proven effective in controlling noise pollution emissions (Rothenberg, 1970; Bergendahl, 1976).

***Neighborhood***

From the 2000s, increasingly more literature dives into the socioeconomic impact of neighborhood noise, from anthropogenic noise to regional sound (Buxtion et al., 2017; Xu et al., 2020; Fan and Weinhold, 2022). Indirect estimations, such as granular-level land use types or population distribution, are used to assess neighborhood noise pollution (Abbaspour et al., 2015). The geographic regions span from the United States (Buxtion et al., 2017) to Europe (Weinhold, 2013; Diaz et al., 2021; Lawlor, 2023), China (Li et al., 2019; Xu et al., 2020), Singapore (Hasegawa and Lau, 2022), Iran (Abbaspour et al., 2015), and Nigeria (Olayinka, 2012).

Different from conventional understanding that neighbourhood noise is a nuance or nonavoidable consequence of modern development, economics literature shows that the impact of neighborhood noise is profound. Le Boennec and Salladarré (2017) show that the noise depreciation in the neighborhood has a significant impact on housing prices. Subjective evaluation factors, which include neighborhood noise, are found better at explaining housing prices than objective measurements (Chasco and Gallo, 2013). Fan and Weinhold (2022) find significant effects of being exposed to neighborhood noise on long-term health outcomes via sleep deprivation. Thus community-level climate actions, such as adopting to renewable energy, implementing urban greening, or using better building design, are effective ways to reduce noise pollution and improve public health (Lawlor, 2023).

***Others***

In addition to the four major sources of noise, other emission origins of noise are also examined in the economics literature, including workplace (Cameron et al., 1972; Dean, 2024), construction site (Ning et al., 2019), vessel noise (Taylor and Mayer, 2023), and wind farms (Jensen et al., 2014; Zou, 2020).

Table 1 summarizes literature under the four major categories of noise sources, as well as outcome variables, which will be analyzed in detail in Section 4.

**Table 1. Summary of Noise Literature with an Economic Approach**

**Panel A: Aircraft**

| Country | Time Period | Outcome Variable | Publication |
|---|---|---|---|
| United States | 1970 - 1972 | Housing | Nelson (1979) |
| Netherlands | 1998 | | Van Praag and Baarsma (2005) |
| Switzerland | 2003 | | Baranzini and Ramirez (2005) |
| Switzerland | 2002 - 2004 | | Boes and Nuesch (2011) |
| Germany | 2006 - 2014 | | Winke (2017) |
| United States | 1990 - 2014 | | Friedt and Cohen (2021b) |
| United States | 2006 - 2017 | | Friedt and Cohen (2021a) |
| Germany | Not Reported | Health | Evans et al. (1998) |
| Germany | Not Reported | | Maschke et al. (2002) |
| United States | 2020 - 2021 | | Jacuzzi et al. (2024) |
| United Kingdom | Not Reported | Sleep | Bugliarello et al. (1976) |
| United States | Not Reported | Shadow prices | Jackson (1971) |
| Netherlands | 1976 | Noise | Alexandre et al. (1980) |
| Sweden | 2005 - 2008 | Neighbourhood composition and households' choice | Lindgren (2021) |
| United States | 1990-2000 | Household income | Ogneva-Himmelberger and Copperman (2010) |
| Germany | 2011-2015 | Violent crime | Hener (2022) |

**Panel B: Railway**

| Country | Time Period | Outcome Variable | Publication |
|---|---|---|---|
| United Kingdom | 1991 | Socioeconomic indicators | Brainard et al. (2004) |
| Singapore<br>Poland | 2005 - 2013<br>2010 - 2015 | Housing | Diao et al. (2016)<br>Trojanek (2023) |

**Panel C: Urban Traffic**

| Country | Time Period | Outcome Variable | Publication |
|---|---|---|---|
| Singapore<br>Singapore<br>Germany | 2006 - 2022<br>2010 - 2018<br>2016 | Housing | Wang et al. (2023)<br>Fan et al. (2021)<br>Kuehnel and Moeckel (2020) |
| Germany<br>United States<br>Not Reported<br>India | 1993/1994<br>1998 - 2007<br>Not Reported<br>2007 | Health | Babisch et al. (2001)<br>Yu et al. (2020)<br>Münzel et al. (2021)<br>Khaiwal et al. (2016) |
| China<br>Ireland<br>Iran<br>India<br>United Kingdom<br>Mexico<br>Brazil | 2019<br>2020<br>Not Reported<br>Not Reported<br>2021<br>2021<br>2012 | Noise | Yang et al. (2020)<br>Basu et al. (2021)<br>Forouhid et al. (2023)<br>Pradeep and Nagendra (2024)<br>Leach et al. (2024)<br>Mora-Quiñones et al. (2024)<br>Fiedler and Zannin (2015) |
| Sweden | 1970 | Investment costs | Bergendahl (1976) |
| Not Reported | Not Reported | Relative price of commodity | Rothenberg (1970) |
| China | 2008 - 2012 | Household-specific utility | Li (2018) |
| Korea | 2017 | Willingness to pay | Kim et al. (2019) |

**Panel D: Neighborhood**

| Country | Time Period | Outcome Variable | Publication |
|---|---|---|---|
| European countries | Not Reported | Noise | Murphy and King (2010) |
| Iran | 2006 | | Abbaspour et al. (2015) |
| United States | Not Reported | | Buxtion et al. (2017) |
| Nigeria | Not Reported | | Olayinka (2012) |
| Brazil | 2012 | | Fiedler and Zannin (2015) |
| Mexico | 2021 | | Mora-Quiñones et al. (2024) |
| China | 2007 | | Shi and Liu (2012) |
| Nigeria | 2005 | | Olayinka (2012) |
| Spain | 2008 | Housing | Chasco and Gallo (2013) |
| Denmark | 2000 - 2011 | | Jensen et al. (2014) |
| France | 2002 - 2008 | | Le Boennec and Salladarré (2017) |
| China | 1991 - 2017 | Marginal utility | Xu et al. (2020) |
| Not Reported | 2021-2022 | Soundscapes and acoustical environments | Hasegawa and Lau (2022) |
| Singapore | 2020 | Resident's satisfaction | Hasegawa and Lau (2022) |
| United States | 1977 to 2019 | Life outcomes of Killer Whales | Taylor and Mayer (2023) |
| European countries | Not Reported | Not Reported | Lawlor (2023) |
| China | 2016 | Health | Zhao et al. (2018) |
| China | 2017 | Mental wellness | Li et al. (2019) |
| Spain | 2020 | Infectious disease incidence | Díaz et al. (2021) |
| European countries (28 countries) | 2003 | Happiness cost | Weinhold (2013) |

**Measure of Noise and Data Source**

***Derived Noise Exposure***

To measure the level of noise pollution, early economics literature starts with using derived noise exposure. A prominent example is proximity to noise origins, such as airports, railways, major roads, and administrative or commercial zones (Rothenberg, 1970; Jackson, 1971; Bergendahl, 1976). The pros and cons of this measure is straightforward: while the distance, either in terms of linear distance or walking distance, is relatively easy to measure, it is not an accurate indicator for noise variation as it can be easily contaminated by other confounding factors. The compromise is possibly due to a shortage of high-frequency or real-time data directly on noise.

A number of economics studies, however, continue with this classical measure, possibly because its convenience outweighs the caveats and the ease in comparison with early literature (Kuehnel and Moeckel, 2020; Wang et al., 2023; Razavi-Termeh et al., 2024). For instance, Kuehnel and Moeckel (2020) estimate a discount of up to 9.6% for particularly loud locations using distance measures. Wang et al. (2023) quantify the price of quietness in the Covid context, showing that the road traffic noise reduces housing rents near major roads by 3.8% immediately after the pandemic outbreak. Similarly, noise pollution declines the prices of housing in the vicinity of wind turbines by 3-7% (Jensen et al., 2014).

***Real-time Monitoring Records***

With technological advancement and the arrival of big-data era, more detailed and real-time noise variations can be captured by sensors or monitors. Economists incorporate data obtained from this source in socioeconomic analysis of noise pollution (Basu et al., 2021; Leach et al., 2024; Jacuzzi et al., 2024). Some of such real-time measurements are used in the laboratories, while others in the general social settings. Zhang et al. (2021) find that exposing rats to structured noise around 65 dB showed significantly impaires hippocampus-related learning and memory. Similar negative association is detected in human being, between 24-hour traffic counts and urinary noradrenaline levels in women aged 30-45 with bedrooms facing streets (Babisch et al., 2001). Trojanek (2023) capitalizes the adverse effect of noise derived from the noise map into housing market, showing that an increase of 1dB above 55dB can induce a 1.8% decline in the apartment value in Poland. Compared with the distance measure, the merit of monitoring record data is clear—it is more accurate and captures the variation in noise only.

However, it relies on specialized devices to collect the noise data; thus it is usually restricted in the laboratories, or if in the general population, the sample size is not large.

***Household Surveys***

Another commonly used method to quantify noise level is household surveys. On the one hand, information on noise exposure can be retrieved from survey questions, such as *"Are you ever confronted with the problems listed below in your home environment?"* (Longitudinal Internet Survey for the Social Sciences, The Netherlands). If respondents agree with *"noise annoyance caused by neighbors"*, it is an indicator for neighbor noise. If they respond to *"noise annoyance caused by factories, traffic or other street sounds"*, an exposure to traffic noise can be quantified. On the other hand, a variety of outcome variables can be extracted from survey questions, such as *"Do you regularly suffer from the following diseases/problems;" "Are you currently taking medicine at least once a week?" or "Has a physician told you this last year that you suffer from the following diseases/problems?"*

For example, Weinhold (2013) collect information on residential noise from the European Quality of Life Survey covering 26,000 respondents across 28 European countries and show that the happiness loss due to noise pollution is estimated as large as172 pounds per household per month. Fan and Weinhold (2022) use the longitudinal household survey in the Netherlands to collect information on exposure to neighbor noise and a batch of health outcomes. Kim et al. (2019) adopts survey data from the South Korea and quantify people's willing to pay to reduce noise annoyance level to zero.

The merit of using household survey data is that the sample size is usually big and can cover a series of potential variables related to noise, socioeconomic consequences, and control variables. Nevertheless, as most household surveys are not specifically designed for noise research, the variation in noise levels may be rough.

***Noise Complaints***

With the advent of administrative data, data on noise complaints derived from the hotlines of government agencies has become another source for economists to conduct noise studies. Residents who suffer from noise annoyance call in local government's hotlines; officers record their basic demographic and socioeconomic information (e.g., gender and neighborhood) and their specific complaints (e.g., *the neighbor upstairs is conducting renovation and is very noisy during the daytime*). Economists then apply the Natural Language Processing method to turn the qualitative complaints into quantitative sentiments, as proxy for the noise level varying across time and geography.

One example using this type of noise complaint data is Fan et al. (2021). Collaborated with local government in Singapore, they combine the quantified noise sentiments with the geographic information of the residents' dwellings and distance to the new bus route, as well as the housing transaction records in the affected region during the sample period. A significant housing price reduction of 3% is discovered, with 1-scale-point increase in noise complaints. The noise pollution from streets almost offsets 18% of the benefit of public transport, calling for noise insulation in dense urban topology.

**Consequence of Noise Pollution**

***Health***

As an important component of human capital, health is among the primary outcomes of noise pollution examined by economists. Bugliarello et al. (1976) pioneer to show that prolonged exposure to sustained and loud noise pollution can lead to hearing loss, with some even being exposed as little as 1 year. Both physical and mental illnesses are detected as outcomes of being exposed to noise pollution. Empirical evidence shows the consequences of noise exposure on cardiovascular disease (Münzel et al., 2021), autoimmune and lung diseases (Fan and Weinhold, 2022), acute and chronic illness (Cameron et al., 1972), and cognitive function (Dean, 2024). In addition, noise can also affect the quality of life via psychophysiological stress (Evans et al., 1998). Significant association is observed between traffic volume and stress hormones (Babisch et al., 2001). Miki et al., (1998) find that compared to quiet condition, an environment with exposure to 90dB white noise increases urinary adrenaline by 76% and cortisol by 105%, with salivary cortisol also significantly higher.

Sleep deprivation is an evident channel through which noise adversely impacts the physical and mental health outcomes (Bugliarello et al., 1976; Gibson and Shrader, 2018; Zou, 2020; Bessone et al., 2021; Fan and Weinhold, 2022). A 1-hour decline in weekly sleep decreases earnings by 1.1% and 5% over the short- and long- run (Gibson and Shrader, 2018). Work productivity is estimated to decrease by 3 percent in a textile training course, associated with a noise increase of 7dB (Dean, 2024). Jacuzzi et al. (2024) estimate that over 74,000 people within the area of aircraft noise exposure were at risk of adverse health effects in the United States.

***Happiness***

In addition to health, happiness is becoming an increasingly powerful tool to measure welfare in the context of noise pollution. It quantifies the utility gain or loss arising from noise pollution. Van Praag and Baarsma (2005) use happiness surveys to value the intangible airport noise in the housing market, as a total of housing stock plus residual shadow costs. Weinhold (2013) estimates the happiness costs of the noise pollution, using survey data across 28 European countries. She finds that residential noise significantly reduces happiness, with the costs as high as 172GBP per household per month. Happiness is a direct measure of the utility. Nevertheless, one potential caveat is that it is usually subject to self-report bias, as different people may have different standard of happiness.

*Violence and Suicides*

On top of the conventionally recognized consequences on health and happiness, noise pollution has even more severe threats to human life. Hener (2022) shows that 4.1 decibels increase in background noise from aircrafts causes a 6.6% increase in the violent crime rate in Germany. Using over 800 wind farm installation events in the United States, Zou (2020) finds robust evidence that the low-frequency noise radiation from wind turbines significantly increase suicide rates in the 15-19 age group and people over 80. Vulnerable population segments disproportionately suffer more from the noise pollution with life-threatening consequences, which sheds light on environmental injustice which will be revisited in Section 4.5.

***Capitalization in Housing Market***

Housing rent or price is another popular market indicator in economics literature used to quantify the cost of noise pollution, by measuring the willingness to pay (Bartik et al., 2019; Kim et al., 2019). Early literature links declining housing prices in the vicinity of six major U.S. airports to the aircraft noise. Nelson (1979) finds that a 5dB increase in aircraft noise would reduce the housing value by around 2.5%. Burgeoning studies from the 2000s contribute to the understanding of how to capitalize noise into the housing market. In the homeowners' market, for instance, Jensen et al. (2014) estimate that noise pollution from wind turbines reduces residential housing price by 3-7% in Denmark's context. A 1.7% price devaluation is estimated for one decibel increase of noise, from the new runway in Germany (Winke, 2017).

In the United States, the noise discount per sale is estimated as high as $25,000. Abatement investments are evaluated to have a 40% return for eligible homes (Friedt and Cohen, 2021b).

The adverse effect of noise pollution is also observed in the renters' market. Noise is estimated to reduce apartment rents by 0.7% per dB, with a larger impact of 1% in airport areas in the Switzerland (Baranzini and Ramirez, 2005). Boes and Nuesch (2011) further show that such rental discount in the Swiss housing market is correlated with unobserved neighborhood factors. Using the outbreak of the Covid as a quasi-natural experiment, Wang et al. (2023) estimate the price of quietness on housing market. They show a decrease by 3.8% of housing rents due to road traffic noise immediately after the outbreak of the pandemic and by 12.7% in the subsequent year.

***Inequality***

Furthermore, noise pollution is related to environmental injustice. Some demographic or socioeconomic groups are more vulnerable to the negative consequences of the noise exposure, such as the younger (15-19 years old) or older (above 80 years old) age groups (Zou, 2020) and lower-income population (Bessone et al., 2021). Ogneva-Himmelberger and Copperman (2010) find that ethnic minority and low-income populations are often exposed to more aircraft noise, as they are selective to live in the vicinity with lower housing prices. Similar evidence is revealed in Lindgren (2021) that renewal of an airport operating contract induced high-income households to move away or choose homes of better quality. Using Indian data, Bessone et al. (2021) show that as low-income adults are more severely affected by noise pollution and deprivation of sleep. The consequence of an increase in sleep duration to compensate indeed entails significant opportunity costs.

In terms of demographics, individuals at the two extreme ends of the age spectrum are more vulnerable to the noise exposure to wind farms, with an increase in the suicide rates for the specific age groups (Zou, 2020). Men, compared to women, are found to have higher noise-induced health risks (Maschke et al., 2002) and higher propensity to be victims of noise-induced physical assaults (Hener, 2022).

***Others***

Other indicators are also used to measure the outcomes of noise exposure, beyond the conventional health, welfare (happiness), violence, housing price/rent, and inequality. Total noise-threshold violation levels and the total system time are examples of the alternatives. Hammad et al. (2017) show that the transport network total system travel time will drop by 32% for each 1dB increase in the noise level. Marginal utility of the regional sound environment quality is also used to quantify the noise consequences along the temporal dimension (Xu et al., 2020). In addition, neighborhood composition is modelled to capture the consequence of noise-induced migration (Lindgren, 2021). Regarding biodiversity, Taylor and Mayer (2023) investigate the significantly lowered fertility and raised the mortality of the Southern Resident Killer whales, as a consequence of the vessel noise pollution from international shipping.

**Economic Approach in Estimating Causal Impact of Noise Pollution**

Early economics literature uses Ordinary Least Squares (OLS) regression analysis to uncover the *correlation* between noise exposure and a bunch of socioeconomic outcomes, as introduced in Section 4 (Bugliarello et al., 1976; Nelson, 1979). Recent studies focus on estimating the *causality*, rather than correlation, of the noise exposure on broader range of outcomes. Economic approaches, such as instrumental variable or Difference-in-Differences estimations are adopted to achieve this goal, as well as using randomized experiments in the laboratory or quasi-natural experiments such as the outbreak of Covid pandemic.

### *Instrumental Variable Estimation*

The key of the instrumental variable estimation in the noise pollution context is to find a valid instrument, which affects the outcomes only through noise exposure. In the econometric terms, it should satisfy the relevance, exogeneity, and exclusion restriction conditions (Wooldridge, 2019).

Nevertheless, finding such an instrument in the noise context is challenging, given the complexity between noise and other environmental factors and potential confounding with unobserved factors. Hener (2022) uses daily variation in aircraft landing approaches due to wind direction to instrument noise levels. A causal effect is identified that with 4.1dB increase in the background noise, the violent crime rate will rise by 6.6%. Fan and Weinhold (2022) exploit neighbor noise as a relatively ex-ante unobservable exogenous shock and use it to instrument for sleep disruption. It is found that the sleep deprivation has significant and adverse effect on health outcomes, such as cardiovascular and auto-immune diseases.

### *Difference-in-Differences Estimation*

Due to the difficulty in identifying a valid instrument, most of recent economics studies use the difference-in-differences method. By constructing treatment vs. control groups and comparing before vs. after shock periods, the DiD method manages to identify the causal impact of noise on a bundle of socioeconomic outcomes. A prominent advantage of the DiD estimation is its clean setup to capture the causal effect, as long as economists can identify an exogeneous shock to construct valid treatment and control groups. However, a potential challenge exists in verifying a crucial pre-trend assumption, which requests a parallel trend between the treatment and control groups before the exogeneous shock. Economists explore various social experiments as quasi-natural experiments to apply the DiD estimation.

In the noise pollution context, for example, Boes and Nüesch (2011) compare apartments in the treatment region—zip codes affected by a new flight regime in the Zurich airport—and those in the control region, to identify the impact of aircraft noise on apartment rents using the DiD strategy. To identify the causal impact of wind turbine noise on suicides, Zou (2020) compare the suicide rates in and outside the vicinity of over 800 wind farms along with their staggered installation events in the United States from 2001 to 2013. Fan et al. (2021) analyze the noise complaints from residents living close to vs. far away from a newly opened bus route in the Singapore context and combine with housing transactions in the DiD setting. They find that bus noise offsets 17.8% of the benefit from convenience of transports.

***Randomized Experiments and Quasi-natural Experiments***

Economics literature generally conducts two types of experiments to capture the plausibly exogeneous shock of noise. The first one is randomized experiments by design. Bessone et al., (2021) design a randomized three-week treatment to urban low-income adults in India, to provide information and improvements to home sleep environments. They thus identify the economic consequences of increasing sleep on cognition, productivity, decision making, and wellbeing. Another example is Dean (2024), which runs two randomized experiments in Kenya by monitoring noise level at workplaces in a textile training course. It is found that with a 7dB increase in noise exposure, the worker productivity declines by 3%.

Other economics studies search for quasi-natural experiments in different socioeconomic settings from various countries. For instance, to capture the impact of aircraft noise on housing markets, Boes and Nüesch (2011) use an unexpected change in flight regulations in the Zurich airport while Winke (2017) the construction of the northwest runway in the airport of Frankfurt, Germany. Fan et al. (2021) adopts the opening of a new bus route as a quasi-natural experiment to identify the impact of public transport on noise sentiments and housing market. Wang et al. (2023) uses the outbreak of the Covid pandemic as a quasi-natural experiment to evaluate the price of quietness, as residents have to stay at home with longer time and expose to noise at a greater extent than pre-Covid period. The DiD estimation is usually coupled with such randomized or quasi-natural experiments to identify the causal impact of noise on housing market or health outcomes.

***Noise Mapping and Soundscape***

Last but not least, noise mapping and soundscape are also used in presenting the noise patterns and tracking the changes. Combining with geographic information, a noise map visualizes the variation in noise levels across time and locations. A soundscape is defined as *"the acoustic environment as perceived or experienced and/or understood by people"* by the International Organization for Standardization (Hasegawa and Lau, 2022a). They are usually adopted in interdisciplinary studies, such as Yang et al. (2020), Hasegawa and Lau (2022a,b), Pradeep and Nagendra (2024).

**Discussion and Conclusion: Noise Pollution, Sustainability, and Household**

Sustainability is a buzz word from the early 2000s. Burgeoning literature investigates the environmental, social, and governance (ESG) factors affecting sustainable growth, mostly from a macro perspective (Murphy and King, 2010; Hasegawa and Lau, 2022a; Lawlor, 2023). There is limited exploration to understand the pathway to achieve sustainability from a household standpoint. Nevertheless, the household is the most granular unit of economic decision-making. It is the forefront of understanding market sentiment regarding environmental shocks, the fundamental unit in addressing environmental challenges, and the essential vehicle for achieving sustainable development.

Much of the public and policymakers' environmental endeavors have been placed on reducing air or water pollution. The ramification of noise pollution has not been fully realized, despite as large as 1.6 million annual healthy life-years loss have been estimated from traffic noise along in Western Europe (WHO, 2011). The adverse consequences of the noise pollution extend beyond physical and mental illness (Cameron et al., 1972; Miki et al., 1998; Bessone et al., 2021; Fan and Weinhold, 2022), affecting crime rates (Zou, 2020; Hener, 2022), the housing market (Nelson, 1979; Jensen et al., 2014; Winke, 2017; Wang et al., 2023), and social welfare and inequality (Van Praag and Baarsma, 2005; Weinhold, 2013; Zou, 2020; Bessone et al., 2021).

This chapter aims to investigate the less understood noise pollution and its consequence on households from an economic perspective. Different from engineer literature focusing on the sound-proof technology and site layout planning (Murphy and King, 2010; Hammad et al., 2017; Hasegawa and Lau, 2022a,b), the economics studies endeavor to isolate *causal* impact of noise pollution on socioeconomic welfare from a lens of households at granular level. Economic approaches are applied to achieve this goal, such as DiD estimation (Boes and Nüesch, 2011; Zou, 2020), Instrumental variable estimation (Fan and Weinhold, 2022; Hener, 2022), randomized experiment (Bessone et al., 2021; Dean, 2024), or quasi-natural experiment (Winke, 2017; Fan et al., 2021; Wang et al., 2023).

To measure the welfare consequences of noise pollution on households and individuals, economics studies combine data on health, housing transactions, crime rates, or happiness with noise measures. To capture temporal variation and nuanced changes in noise levels, economists apply distance variations to the origin of noise (Jensen et al., 2014; Kuehnel and Moeckel, 2020; Wang et al., 2023; Razavi-Termeh et al., 2024), adopt real-time noise monitoring measure (Babisch et al., 2001; Trojanek, 2023; Jacuzzi et al., 2024), extract information from household surveys (Weinhold, 2013; Kim et al., 2019), or use the Natural Language Processing method to analyze noise complaints (Fan et al., 2021).

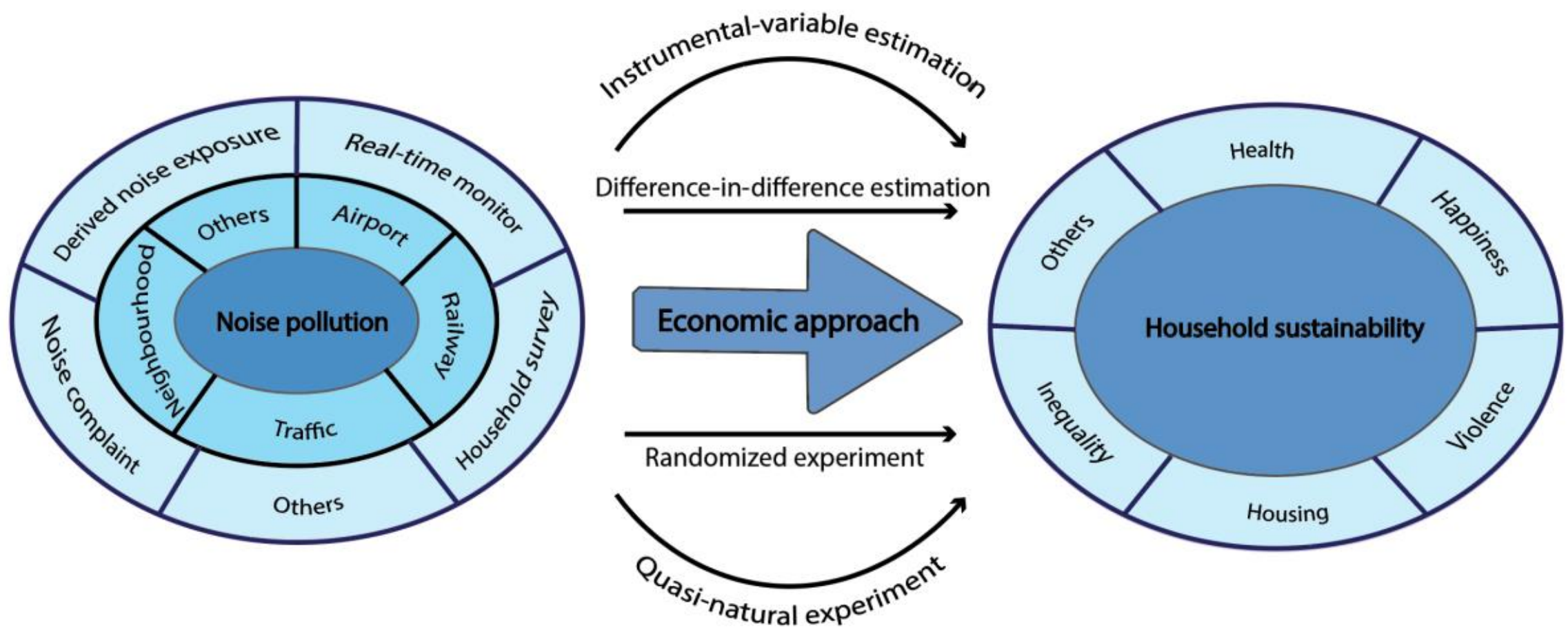

**Figure 1. Noise Pollution and Household Sustainability in an Economic Framework**

Figure 1 summarizes the relationship between noise pollution and household sustainability in an economic framework. The magnitude of the adverse effect of noise pollution on households is striking. Hearing loss can occur with as little as 1-year exposure to sustainable and loud noise (Bugliarello et al., 1976). A 7 dB increase in noise can induce 3% decline in work productivity (Dean, 2024) and environmental noise above 90 dB can double the cortisol level (Miki et al., 1998). Moreover, merely 4 dB increase in aircraft noise can cause 6.6% increase in violence rates in the vicinity (Hener, 2022). The housing market has sensitive momentum with the noise exposure, estimated at a discount as high as $25,000 per sale in the United States (Friedt and Cohen, 2021b).

The causal effects of noise pollution identified by economics studies can guide cost-benefit analysis and direct policy efforts in the abatement investment. Rothenberg (1970) advocates for congestion charges, without which pollution would be worsen due to imbalanced investments in public and private measures. Bergendahl (1976) promotes vehicle taxes, which are collected under the "Polluter-Pays Principle" in Sweden to reduce noise pollution emissions and leads to the development of quieter vehicles. Friedt and Cohen (2021b) estimate a 40% return to abatement investments for eligible homes. Low Traffic Neighborhoods are introduced in the United Kingdom, with an evidently significant reduction in noise (Leach et al., 2024). Hammer et al. (2018) set up an I-ACT framework to analyze noise pollution policies across different countries and identify four key drivers of change in noise pollution, which are information systems, public awareness, leadership and coordination, and tools. The policy efforts continue to extend to developing countries, with designed experiments in India and Kenya to identify the adverse effects from noise exposure (Bessone et al. 2021; Dean, 2024). Future studies are warranted along this line to understand the socioeconomic consequences of noise pollution in general population in the less developed regions/nations. Environmental justice also calls for further investigation on policies and actions to help noise vulnerable population, who disproportionately suffer more from the noise pollution.

**Further Reading**

**Notes:**

---

[i]Email: yi.fan@nus.edu.sg. I thank Isabell Chew Pei Ling, Kangyi Zhang, and Xiaoqian Zhang for excellent research support. I acknowledge financial support from the National University of Singapore Academic Research Fund A-8002711-00-00.